\documentclass[
 reprint,
]{revtex4-1}

\usepackage{graphicx}
\usepackage{dcolumn}
\usepackage{mathrsfs}
\usepackage{amsmath}
\usepackage{bm}
\newtheorem{theorem}{Theorem}

\begin{document}

\preprint{APS/123-QED}

\title{Analytical Results of k-core Pruning Process on Multi-layer Networks}

\author{Rui-jie Wu}
\author{Yi-Xiu Kong}
\author{Gui-Yuan Shi}
\email{guiyuan.shi@unifr.ch}
\author{Yi-Cheng Zhang}
 
\affiliation{Department of Physics, University of Fribourg, Fribourg 1700, Switzerland}

\date{\today}

\begin{abstract}

Multi-layer networks or multiplex networks are generally considered as the networks that have the same set of vertices but different types of edges. Multi-layer networks are especially useful when describing the systems with several kinds of interactions. In this paper we study the analytical solution of $\textbf{k}$-core pruning process on multi-layer networks. $k$-core decomposition is a widely used method to find the dense core of the network. Previously the Nonbacktracking Expand Branch (NBEB) is found to be able to easily derive the exact analytical results in the $k$-core pruning process. Here we further extend this method to solve the $\textbf{k}$-core pruning process on multi-layer networks by designing a variation of the method called Multicolor Nonbacktracking Expand Branch (MNEB). Our results show that, given any initial multi-layer network, Multicolor Nonbacktracking Expand Branch can offer the exact solution for each intermediate state of the pruning process, these results do not only apply to uncorrelated network, but also apply to networks with either interlayer correlations or in-layer correlations.

\end{abstract}

\pacs{Valid PACS appear here}
\maketitle

\section{Introduction}

Graphs are often used to model the systems that consist of interacting people or entities, where the vertices represent people or entities and the edges represent connections. Nowadays many graphs are built this way, from a variety of systems and applications, such as online social networks, e-commerce platform and even protein interaction networks. One of the most important tasks in analyzing these graphs is to find the densest part of the network where the vertices are closely related to each other~\cite{du2007community,fortunato2010community,papadopoulos2012community,weng2013virality}. The most commonly used algorithm for this problem is called the k-core decomposition, in which the goal is to find the subgraph consists of the vertices that are left after all vertices whose degrees less than $k$ have been removed. $k$-core decomposition is widely used to help visualize network structures~\cite{alvarez2006large,serrano2009extracting}, understand and explain the collaborative process in social networks~\cite{goltsev2006k,dorogovtsev2006k}, describe protein functions based on protein-protein networks~\cite{altaf2006development,li2010computational}, and promote network methods for large text summaries~\cite{antiqueira2009complex}, and so on. 

Previously, many researchers~\cite{dorogovtsev2006k,fernholz2004cores,schwarz2006onset,baxter2015critical} have focused on solving the $k$-core decomposition problem on single layer network. The analytical results on the final size as well as the structure of $k$-core on large uncorrelated networks have been obtained by Baxter et al~\cite{baxter2015critical}. Based on this theoretical framework, Shi et al.~\cite{shi2018analytical} further find the analytical results on the intermediate states of the pruning process that depict the entire critical phenomenon. In addition, Wu et al.~\cite{wu2018using} show that the Nonbacktracking Expand Branch proposed by et al.~\cite{timar2017nonbacktracking} can be used to obtain the exact results of $k$-core pruning process on correlated networks. The Nonbacktracking Expand Branch is an alternative representation to the usual adjacency matrix of a network, it is constructed as an infinite tree having the same local structural information with the given network, when observed by nonbacktracking walkers. 

These findings are important to our understanding of the structure of the complex networks, and the results on correlated networks shed lights on the possibility of analyzing the realistic networks with a theoretical approach. 
On the other hand, in real-world scenarios, it is common that we have to deal with systems that consist of many different types of interactions. As a result, the systems cannot be represented by a single layer network. For systems with multiple kinds of connections, we naturally use multi-layer networks (also called as multiplex networks, multidimensional networks, etc.)~\cite{boccaletti2014structure} that have the same set of vertices but different kinds of edges to represent these systems. In a multi-layer network, each kind of connection is represented by a unique layer, and the same vertex is allowed to have different network structures in different layers. Fig.~\ref{multilayer-fig1} \textbf{(a-c)} show a simple example of multi-layer network.

Here we show that by assigning different colors to distinguish the types of interactions, we can use the Multicolor NonBacktracking Expand Branch (MNEB) method to analytically obtain the results of each step in the $\textbf{k}$-core pruning process in a multi-layer network. It is worth noting that our method is not limited to the analysis of uncorrelated networks, it also works for correlated networks as a natural extension.

\section{Method}

$\textbf{k}$-core decomposition on multi-layer networks is to find the largest subgraph in which the degree of each vertex is at least $k_i$ in the $i^{th}$ layer (here $\textbf{k}$ is a non-negative integral vector). In the previous paper~\cite{azimi2014k}, the researchers give the analytical result of the final size of $\textbf{k}$-core on multi-layer networks. Here we show that the NonBacktracking Expansion Branching (NBEB) method can be used to obtain the complete solution of $\textbf{k}$-core decomposition on the multi-layer, in which not only the final state but each intermediate state of the pruning process can be obtained analytically. 


Given a multi-layer network in which each layer is a simple graph, the standard pruning algorithm for $\textbf{k}$-core decomposition is: for a given sequence of $k_i$, at each step, we remove the vertices that have degrees less than $k_i$ in $i^{th}$ layer network. In the following we analyze in detail the pruning process, and attempt to give the size of the remaining vertices after each step.



First of all, let us introduce the definitions of the terms that will be used in the following of the paper. Suppose a given multi-layer network has $R$ layers. For convenience, we assign each layer with a color $c_i$ to distinguish the edges that belong to different network layers. A 'stub' is defined as a combination of an edge $e_{[i]}$ and one of its end vertex $V$, denoted by $(e_{[i]},V)$, where the subscript $[i]$ here means that it belongs to the $i^{th}$ layer, and $i$ can be any integer in $[1,R]$. Obviously, the stub $(e_{[i]},V)$ has the color $c_i$. We denote by $j_i$ the degree of vertex $V$  in the $i^{th}$ layer. In the $i^{th}$ layer, we define the neighbor stubs set of vertex $V$: $S_i(V)=\{(e_{[i],1},V_{[i],1}),(e_{[i],2},V_{[i],2}),\dots (e_{[i],j_i},V_{[i],j_i})\}$, here $\{V_{[i],1},V_{[i],2}, \dots, V_{[i],j_i}\}$ are the neighbors of $V$ in the $i^{th}$ layer, and $\{e_{[i],1},e_{[i],2}, \dots, e_{[i],j_i}\}$ are the edges connecting them to $V$, and the excess neighbor stubs set of any stub $(e_{[i]},V)$ in $i^{th}$ layer to be the complementary set of $\{(e_{[i]},V^*)\}$ in $S_i(V)$, which is $S_i(e_{[i]},V)=S_i(V)/\{(e_{[i]},V^*)\}$($V^*$ is the neighbor of V via $e_{[i]}$).

Consequently, for the whole multi-layer network, we define the neighbor stubs set of vertex $V$:
\begin{equation}
S(V)=\bigcup_{j=1}^{R}S_j(V),
\end{equation}
and excess neighbor stubs set of $(e_{[i]},V)$:
\begin{equation}
S(e_{[i]},V)=\big{(}\bigcup_{j=1,j\ne i}^{R}S_j(V)\big{)}\cup S_i(e_{[i]},V).
\end{equation}
Note that the above expression of excess neighbor stubs set $S(e_{[i]},V)$ is equivalent to the complementary set of $\{(e_{[i]},V^*)\}$ in $S(V)$.

\begin{figure}[!h]
\centering
\includegraphics[width=8cm]{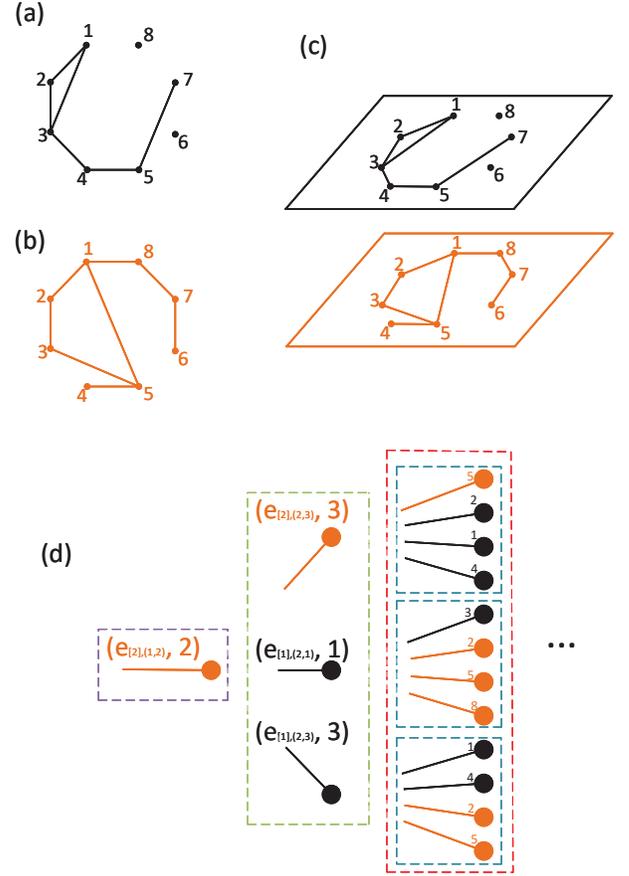}
\caption{
An exemplary a two-layer network. \textbf{(a)} The first layer; \textbf{(b)} The second layer; \textbf{(c)} The two-layer network consists of the two layers in \textbf{(a)} and \textbf{(b)}; \textbf{(d)} The Multicolor Nonbacktracking Expand Branch $B(e_{[2],(1,2)}, 2)$of the two-layer network shown in \textbf{(b)}. Here $e_{[h],(i,j)}$ denotes the edge that connects vertex $i$ and vertex $j$ in the $h^{th}$ layer. The purple, green and red boxes represents the first, second and third stratum of the MNEB, respectively. Note that the excess neighbor stubs set $S(e_{[2],(1,2)}, 2)=\{(e_{[2],(2,3)}, 3),(e_{[1],(2,1)}, 1),(e_{[1],(2,3)}, 3)\}$. The child stubs of stub $(e_{[2],(1,2)}, 2)$ are all the elements in $S(e_{[2],(1,2)}, 2)$, as shown in the green box. The $3$ blue boxes represent the child stubs of the $3$ corresponding stubs in the second stratum, respectively.
}
\label{multilayer-fig1}
\end{figure}

\begin{figure}[!h]
\centering
\includegraphics[width=8cm]{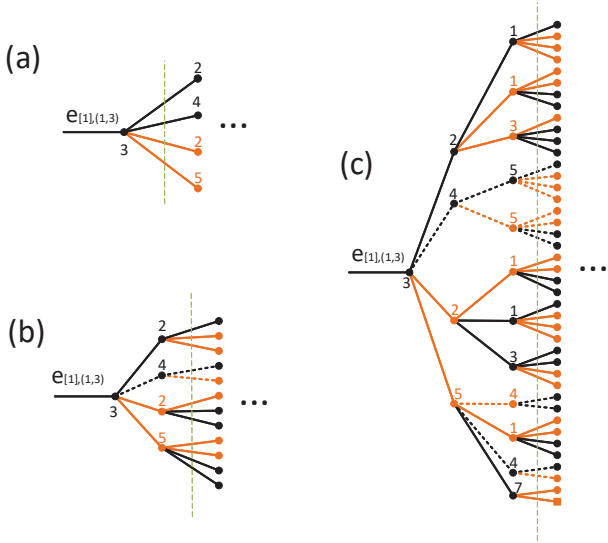}
\caption{
A graphic illustration of $Y_{[1],n}$ of the 2-layer network shown in Fig.~\ref{multilayer-fig1}. The MNEB $B(e_{[1],3},3)$ has the root colored with $c_1=black$. For example, we perfrom $\mathbf{k}=(1,2)$-core decomposition on the network. The green dashed line is the indication line of the first $n$ strata of the MNEB. The solid lines in the MNEB represents the stubs that fulfill the condition $2$ under given $\textbf{k}$. The condition that an MNEB belongs to $Y_{[1],n}$ is then decided by whether the MNEB has a solid subbranch crossing the green dashed indication line. 
\textbf{(a)} shows that $B(e_{[1],3},3)\in Y_{[1],1}$. (b) shows that $B(e_{[1],(1,3)},3)\in Y_{[1],2}$. (c) shows that $B(e_{[1],(1,3)},3)\in Y_{[1],3}$. In addition, for  $\mathbf{k}=(2,2)$, we can also see that $B(e_{[1],(1,3)},3)\in Y_{[1],1}$ from \textbf{(a)}, $B(e_{[1],(1,3)},3)\in Y_{[1],2}$ from \textbf{(b)}. But for $n=3$, we can not find such a subbranch that for each black vertex in the first 3 strata of this subbranch, the number of red child vertices are no less than $2$, and the number of black child vertices are no less than $1$, and for each red vertex in the first $3$ strata of this subbranch,  the number of red child vertices are no less than $1$, and the number of black child vertices are no less than $2$. Therefore, $B(e_{[1],(1,3)},3)\notin Y_{[1],3}$. Here the discs represent the vertices that have child vertices, and the squares represent the vertices without child vertices.
}
\label{multilayer-fig2}
\end{figure}


Similar to the definition of NonBacktracking Expansion Branch (NBEB) in one layer network~\cite{wu2018using}, starting from any stub $(e_{[i]}, V)$, we can define such a tree-like structure which we call the Multicolor Nonbacktracking Expansion Branch(MNEB). 
The chosen stub $(e_{[i]}, V)$(with color $c_i$) is the root of the MNEB, regarded as the first stratum. For any known $n^{th}$ stratum of the MNEB, we can further find the child stubs of each stub in the $n^{th}$ stratum that are all the elements in its excess neighbor stubs set, and all these child stubs constitute the $(n + 1)^{th}$ stratum of the MNEB. For example, if we have a stub $(e_{[i]}^*, V^*)$ in the $n^{th}$ stratum, its child stubs are all the stubs that belong to $S(e_{[i]}^*, V^*)$. 
We can continue this process so that we obtain the MNEB of the stub $(e_{[i]}, V )$, denoted by $B(e_{[i]}, V )$. Obviously, there can be different colored stubs in one MNEB. Fig.~\ref{multilayer-fig1} \textbf{(d)} gives an illustration of how the MNEB is constructed. 

For a given $R$-dimensional positive integral vector $\textbf{k}=(k_1,k_2,\dots,k_R)$, we can find a set of MNEBs $Y_{[i],n}$ ($n$ is a positive integer) for each $1\leq i \leq R$ that meet the following two conditions: 1. The root of the MNEB is colored with $c_i$. 2. there exists a subbranch of the MNEB that contains the root stub, for each vertex colored with $c_j$ in the first $n$ layers of this subbranch, it has at least $k_j-1$ $c_j$ colored child vertices, and at least $k_l$ colored child vertices for every $l\ne j$($1\leq l \leq R$). Fig.~\ref{multilayer-fig2} show the details of how to decide whether an MNEB belongs to $Y_{[i],n}$. When $n=0$, for each $1\leq i \leq R$, we define $Y_{[i],0}$ as all the MNEBs whose  roots are colored with $c_i$. Obviously, $Y_{[i],0}\supset Y_{[i],1}\supset \dots \supset Y_{[i],n}\supset \dots$. We denote $S_{MNEB}(V)$ to be the set of MNEBs of all the stubs in $S(V)$, and $S_{MNEB}(e_{[i]},V)$ to be the set of MNEBs of all the stubs in $S(e_{[i]},V)$. It is easy to obtain the following theorem from the definition of $Y_{[i],n}$:

\begin{theorem}
For a stub $(e_{[i]},V)$ in the $i^{th}$ layer, the MNEB $B(e_{[i]},V)$ belongs to $Y_{[i],n}$, if and only if among all the MNEBs in $S_{MNEB}(e_{[i]},V)$, at least $k_i-1$ MNEBs belong to $Y_{[i],n-1}$, and at least $k_l$ MNEBs belong to $Y_{[l],n-1}$ for every $l\ne j$($1\leq l \leq R$).
\end{theorem}

Let $S_n$ be the set of the remaining vertices after $n^{th}$ pruning, and the following theorem can be established,

\begin{theorem}
Denote by $V$ a vertices in the network, $V\in S_n$, if and only if among the MNEBs in $S_{MNEB}(V)$, for every $1\leq l \leq R$, at least $k_l$ MNEBs belong to $Y_{[l],n-1}$.
\end{theorem}

The proof of Theorem 2 is given in Appendix. An illustration of the $5$ MNEBs in $S_{MNEB}(1)$ of the network from Fig.~\ref{multilayer-fig1} along with a short exemplary analysis using Theorem $2$ are presented in Appendix as well.

\section{Analysis on large uncorrelated multi-layer networks}
As a special case, we start with the partly uncorrelated multi-layer networks, in which the degrees of vertices are uncorrelated in each layer but the degrees of vertices in different layers are allowed to be interdependent. 

For a random vertex $V$, it has the degree serie $(i_1,i_2,\dots,i_R)$ on a large uncorrelated multi-layer network, where $i_1,i_2,\dots,i_R$ represent the degrees of a vertex in $1,2,\dots,R$ layer of the network respectively. The joint degree distribution probability of the vertex is denoted by $p_{i_1,i_2,\dots,i_R}$, and the joint excess degree distribution of the vertex in the $j^{th}$ layer is denoted by $q^{[j]}_{i_1,i_2,\dots,i_R}$, which is the probability that following a randomly chosen edge in the $j^{th}$ layer and one of its endpoint has the excess degree $j_i$, while in the $h^{th}$ layer($h\ne j$), its degree is $i_h$. After that we can define the following two generating functions:

\begin{equation}
G_0(z_1,z_2,\dots,z_R)=\sum_{i_1=0}^{\infty}\dots\sum_{i_R=0}^{\infty}p_{i_1,i_2,\dots,i_R}z_1^{i_1}\cdot\dots\cdot z_R^{i_R}
\end{equation}  

\begin{equation}
G^{[j]}_1(z_1,z_2,\dots,z_R)=\sum_{i_1=0}^{\infty}\dots\sum_{i_R=0}^{\infty}q^{[j]}_{i_1,i_2,\dots,i_R}z_1^{i_1}\cdot\dots\cdot z_R^{i_R},
\end{equation}  
where the superscript in the second definition indicates the generating function is defined in the $j^{th}$ layer. 

These two generating functions are related by:
\begin{equation}
G^{[j]}_1(z_1,z_2,\dots,z_R)=\frac{1}{c_j}\frac{\partial G_0(z_1,z_2,\dots,z_R)}{\partial z_j},
\end{equation} 
where $c_j$ is the average degree of the $j^{th}$ layer network. 
For convenience, we introduce the following denotation:
\begin{equation*}
\sum_{\textbf{i}=\textbf{x}}^{\textbf{t}}=\sum_{i_1=x_1}^{t_1}\sum_{i_2=x_2}^{t_2}\dots\sum_{i_R=x_R}^{t_R},
\end{equation*}
$\textbf{x}$ and $\textbf{t}$ are two fixed integral $R$ dimensional vectors. The above denotion means to take the sum for $\textbf{i}$ from the first component to the last component. Of course there must be $t_j\geq x_j$ for each $1\leq j \leq R$.

Let $y_{[j],n}$ to be the probability that an MNEB whose root is colored with $c_j$ belongs to $Y_{[j],n}$, then from theorem 1 we can obtain the recursive relationship:

\begin{align}
y_{[j],n}&=\sum_{\textbf{i}=\textbf{k}_j}^{\infty}q^{[j]}_{\textbf{i}}\sum_{\textbf{m}=\textbf{k}_j}^{\textbf{i}}\prod_{l=1}^{R}{i_l \choose m_l}(y_{[l],n-1})^{m_l}(1-y_{[l],n-1})^{i_l-m_l} \nonumber\\
            &=\sum_{\textbf{m}=\textbf{k}_j}^{\infty}\frac{\partial^{(m_1+\dots+m_R)}G^{[j]}_1}{\partial z_1^{m_1}\dots \partial z_R^{m_R}}\bigg{|}_{\textbf{z}=\textbf{1}-\textbf{y}_{n-1}}\prod_{l=1}^{R}\frac{(y_{[l],n-1})^{m_l}}{m_l!},\label{yn}
\end{align}
where $y_{[j],0}=1$ for every $1\leq j \leq R$. In the above equation, note that the summation indexes $\textbf{i}=(i_1,i_2\dots i_R)$, $\textbf{m}=(m_1, m_2\dots m_R)$ are vectors. When performing a $\textbf{k}$-core on the multi-layer network($\textbf{k}$ is a vector), we define an $R$-dimensional integral vector $\textbf{k}_j=(k_1,\dots,k_{j-1}, k_j-1, k_{j+1}\dots k_R)$. $\textbf{y}_{n-1}$ denotes the $R$-dimensional vector $(y_{[1],n-1},y_{[2],n-1} \dots y_{[R],n-1})$. Therefore $\textbf{1}-\textbf{y}_{n-1}$ is the $R$- dinmensional vector $(1-y_{[1],n-1},1-y_{[2],n-1}, \dots 1-y_{[R],n-1})$. 

Then we denote by $s_n$ the probability that a randomly chosen vetex belongs to $S_n$.

\begin{align}
s_n&=\sum_{\textbf{i}=\textbf{k}}^{\infty}p_{\textbf{i}}\sum_{\textbf{m}=\textbf{k}}^{\textbf{i}}\prod_{l=1}^{R}{i_l \choose m_l}(y_{[l],n-1})^{m_l}(1-y_{[l],n-1})^{i_l-m_l} \nonumber\\
            &=\sum_{\textbf{m}=\textbf{k}}^{\infty}\frac{\partial^{(m_1+\dots+m_R)}G_0}{\partial z_1^{m_1}\dots \partial z_R^{m_R}}\bigg{|}_{\textbf{z}=\textbf{1}-\textbf{y}_{n-1}}\prod_{l=1}^{R}\frac{(y_{[l],n-1})^{m_l}}{m_l!},\label{sn}
\end{align}

In the complete uncorrelated multi-layer network, which there exists no correlation in different layers, we have $p_{i_1,\dots,i_R}=p_{i_1}\cdot p_{i_2}\cdot \dots \cdot p_{i_R}$, hence the generating function can be simplified:
\begin{equation}
G_0(z_1,z_2,\dots,z_R)=\prod_{j=1}^{R}G_{[j],0}(z_j),
\end{equation}
\begin{equation}
G^{[j]}_1(z_1,z_2,\dots,z_R)=G_{[j],1}(z_j)\prod_{h=1,h\ne j}^{R}G_{[h],0}(z_h).
\end{equation}
Here $G_{[j],0}(z_j)$ and $G_{[j],1}(z_j)$ are the generating function of degree distribution and excess degree distribution of the $j^{th}$ layer network, respectively.
Take these two genereting functions into Eq.~\ref{yn} and Eq.~\ref{sn},
\begin{align}
y_{[j],n}&=\left[1-\sum_{m=0}^{k_j-2}\frac{{(y}_{[j],n-1}{)}^m}{m!}G_{[j],1}^{(m)}(1-y_{[j],n-1})\right] \nonumber\\
&\times\prod_{h=1,h\ne j}^{R}\left[1-\sum_{m=0}^{k_h-1}\frac{{(y}_{[h],n-1}{)}^m}{m!}G_{[h],0}^{(m)}(1-y_{[h],n-1})\right],
\end{align}
and:
\begin{equation}
s_n=\prod_{h=1}^{R}\left[1-\sum_{m=0}^{k_h-1}\frac{{(y_{[h],n-1})}^m}{m!}G_{[h],0}^{(m)}(1-y_{[h],n-1})\right].
\end{equation}
Below we further perform several numerical simulations to validate the theoretical results in section~\ref{numerical}.

\section{Analysis on large correlated multi-layer networks}

Next we study the general case that $\textbf{k}$-core decomposition performed on large correlated networks. We use similar notation with M. Newman~\cite{newman2002assortative}, define $e^{[h]}_{\textbf{i},\textbf{j}}$ as the probability that upon following a randomly chosen edge in the $h^{th}$ layer network, for its two endpoint $V$ snd $V^*$, the excess degree of $V$ being $i_h$ in the $h^{th}$ layer, the degree of $V$ in any other layer $l$ being $i_l$, meanwhile the the excess degree of $V^*$ being $j_h$ in the $h^{th}$ layer, the degree of $V^*$ in any other layer $l$ being $j_l$.

The subscripts of the probability $e^{[h]}_{\textbf{i},\textbf{j}}$, $\textbf{i}=(i_1,i_2,\dots,i_R)$ and $\textbf{j}=(j_1,j_2,\dots,j_R)$ are two non-negative integral vectors. 

$e^{[h]}_{\textbf{i},\textbf{j}}$ must meet the following two conditions:
\begin{equation}
e^{[h]}_{\textbf{i},\textbf{j}}=e^{[h]}_{\textbf{j},\textbf{i}},
\end{equation}
and
\begin{equation}
\sum_{\textbf{j}=\textbf{0}}^{\infty}e^{[h]}_{\textbf{i},\textbf{j}}=q_{\textbf{i}}^{[h]}.
\end{equation}

Then we can denote by $y_{[h],\textbf{i},n}$ the probability that an MNEB $B(e_{[h]},V)$ belongs to $Y_{[h],n}$, given that the other endvertex of $e_{[h]}$(denoted by $V^*$), has excess degree in the $h^{th}$ layer equals to $i_h$ and its degree in the $l^{th}$ layer equals to $i_l$ for every $l\ne h$($1\leq l \leq R$). If $q^{[h]}_{\textbf{i}}=0$, which means $y_{[h],\textbf{i},n}$ does not exists, we can define $y_{[h],\textbf{i},n}=1$. Obviously $y_{[h],\textbf{i},0}=1$. From theorem 1, we have the following recursive relationship:
\begin{align}
q^{[h]}_{\textbf{i}}\cdot y_{[h],\textbf{i},n}&=\sum_{\textbf{j}=\textbf{k}_h}^{\infty}e^{[h]}_{\textbf{i},\textbf{j}}\sum_{\textbf{m}=\textbf{k}_h}^{\textbf{i}}\prod_{l=1}^{R}\bigg{[}{j_l \choose m_l}(y_{[l],\textbf{j}_l^{[h]},n-1})^{m_l} \nonumber\\
&\times(1-y_{[l],\textbf{j}_l^{[h]},n-1})^{i_l-m_l}\bigg{]} ,
\end{align}
here:
\begin{displaymath}
\textbf{j}_l^{[h]} = \left\{ \begin{array}{ll}
(j_1,j_2,\dots,j_R) & \textrm{if $l=h$}\\
(j_1,\dots, j_l-1,\dots, j_h+1,\dots, j_R) & \textrm{if $l<h$}\\
(j_1,,\dots, j_h+1\dots, j_l-1,\dots, j_R) & \textrm{if $l>h$},
\end{array} \right.
\end{displaymath}
and from theorem 2:
\begin{equation}
s_n=\sum_{\textbf{j}=\textbf{k}}^{\infty}p_{\textbf{j}}\sum_{\textbf{m}=\textbf{k}}^{\textbf{j}}\prod_{l=1}^{R}{j_l \choose m_l}(y_{[l],\textbf{j}_l,n-1})^{m_l}(1-y_{[l],\textbf{j}_l,n-1})^{j_l-m_l}, 
\end{equation}
where $\textbf{j}_l$ denotes $(j_1,\dots,j_l-1,\dots,j_R)$.

When there are no correlation in the multi-layer network, that means $e^{[h]}_{\textbf{i},\textbf{j}}=q^{[h]}_{\textbf{i}}\cdot q^{[h]}_{\textbf{j}}$. The results can be easily found to be consistent with the previous results on large uncorrelated multi-layer networks.

\section{Numerical simulations} \label{numerical}

\begin{figure}[!h]
\centering
\includegraphics[width=8cm]{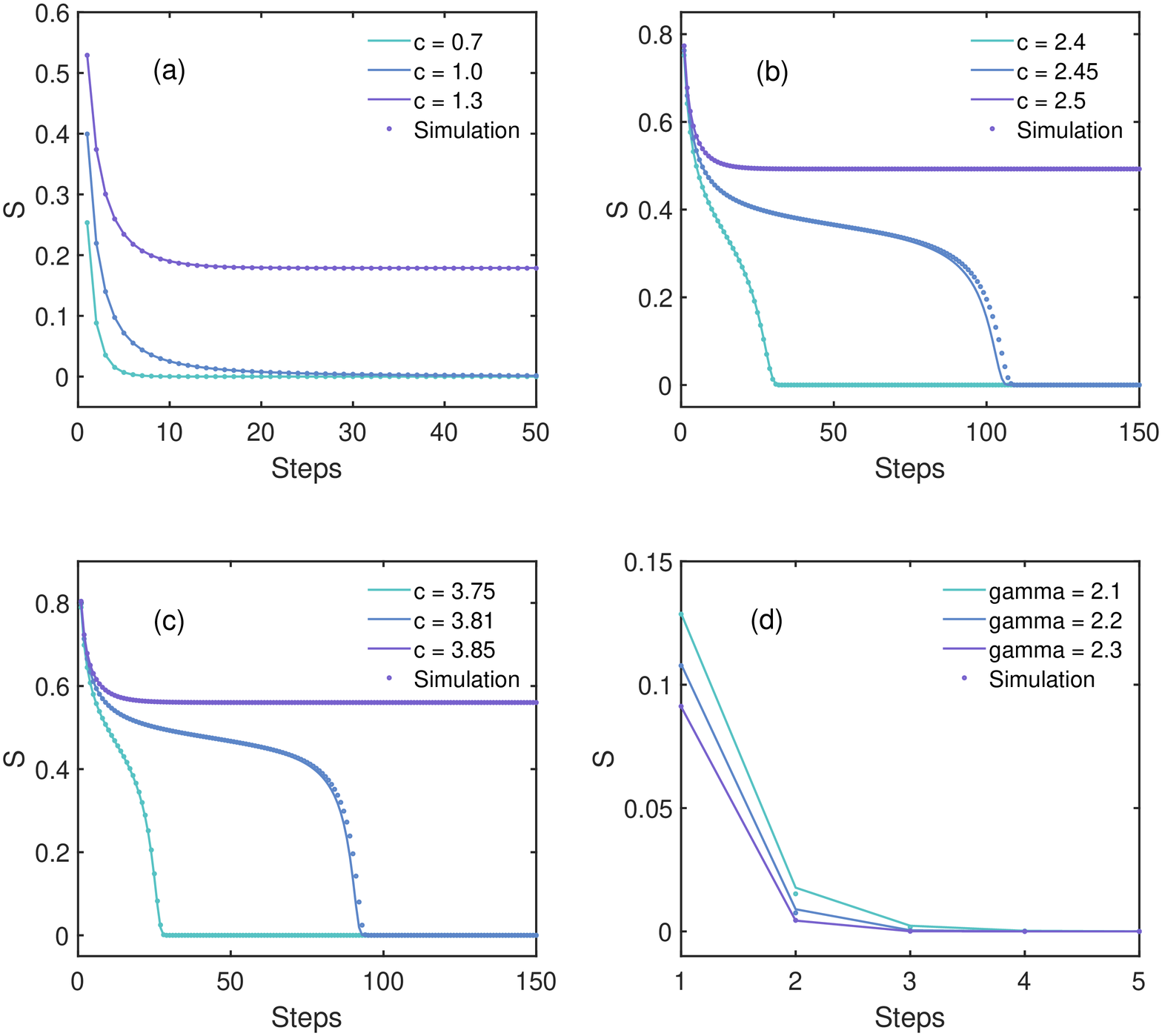}
\caption{
\textbf{
Theoretical results (solid lines) and numerical simulation results (dots) for $\textbf{k}$-core decompositions on multi-layer networks with identical parameters.} 
All simulations are performed on networks with $10^7$ vertices, except that in the simulations of $c=2.45$ in \textbf{(b)}, $c=3.81$ in \textbf{(c)} and $\gamma=2.1$ in \textbf{(d)}, the networks contain $5\times10^7$ vertices. \textbf{(a)} shows the results of $(1,1)$-core pruning process on 3 different two-layer uncorrelated ER networks. Note that in this case it shows a continuous phase transition. \textbf{(b)} shows the results of $(1,1,1)$-core pruning process on 3 different three-layer uncorrelated ER networks. In this case the networks exhibit a discontinuous phase transition. 
\textbf{(c)} shows the results of $(2,2)$-core pruning process on 3 different two-layer uncorrelated ER networks. Note that In this case the networks exhibit a discontinuous phase transition, different from the case shown in \textbf{a}. 
\textbf{(d)} shows the results of $(2,2)$-core pruning process on 3 different two-layer uncorrelated SF networks. In this case the $(2,2)$-core does not exist for $\gamma > 2$.
}
\label{multilayer-fig4}
\end{figure}

To validate our method, we perform several $\textbf{k}$-core decompositions on complete uncorrelated multi-layer Erd\H{o}s-R\'enyi networks (ER networks) and Scale-Free networks (SF networks). The results are shown in Fig.~\ref{multilayer-fig4}. It can be seen that our theoretical results are in perfect accordance with the numerical simulations.

\section{Conclusion}

Overall, in this paper we derive a new variation of the Nonbacktracking Expand Branch called the Multicolor Nonbacktracking Expand Branch specially designed to solve the $\textbf{k}$-core pruning process on Multi-layer networks. In a multi-layer network, each layer of the network is assigned with a unique color, then Multicolor Nonbacktracking Expand Branch is constructed as an infinite tree having the same local structural information with the given multi-layer network, when observed by nonbacktracking walkers. We find that with this representation, one can easily obtain the analytical results of $\textbf{k}$-core pruning process on any given multi-layer network, regardless the correlation exists or not. The theoretical results obtained by our method are further validated by numerical simulations. Our method opens new possibilities to analytically solve the $\textbf{k}$-core pruning process on any given multi-layer network, which is valuable for both theoretical studies and real-world applications.

\section{Appendix}

\noindent
\textbf{Proof of Theorem 2}:\\
We use mathematical induction to prove the theorem. It is obvious that the theorem holds for $n = 1$. Now we prove that if the theorem is true for $n - 1$, the theorem can be established for $n$. 

Firstly we prove the sufficiency, that is, for every $1\leq l \leq R$, when at least $k_l$ MNEBs in $S_{MNEB}(V)$ belong to $Y_{[l],n-1}$, there must be $V\in S_n$. Since for every $1\leq l \leq R$, $Y_{[l],n-1}\subset Y_{[l],n-2}$, we obtain $V\in S_{n-1}$, and in any given layer(for instance, the $i^{th}$ layer), suppose that $\{B(e_{[i],j_1},V_{j_1}),\dots,B(e_{[i],j_m},V_{j_m})\}\subset  S_{MNEB}(V)$ belong to $Y_{[i],n-1}$, here $m \geq k_i$, so for each $1\leq r\leq m$, in $S_{MNEB}(e_{[i],j_r},V_{j_r})$, at least $k_i-1$ MNEBs belong to $Y_{[i],n-2}$, and at least $k_l$ MNEBs belong to $Y_{[l],n-2}$ for every $l\ne j$($1\leq l \leq R$). On the other hand, $B(e_{[i],j_r},V)\in Y_{[i],n}\subset Y_{[i],n-2}$, so in $S_{MNEB}(V_{j_r})$, for every $1\leq l \leq R$, at least $k_l$ MNEBs belong to $Y_{[l],n-2}$. The induction hypothesis gives $V_{j_r}\in S_{n-1}$. Therefore, in the $(n-1)^{th}$ pruning, in any given $i^{th}$ layer, at least $k_i$ neighbors of $V$ are retained. We can conclude that $V$ is still retained in the $n^{th}$ pruning.

Next we prove the necessity. We attempt to prove that when there exists an $l$ in $[1,R]$, that satisfies that at most $k_l-1$ MNEBs in $S_{MNEB}(V)$ belong to $Y_{[l],n-1}$, there must be $V\notin S_n$. Since for an MNEB $B(e_{[l],r},V_r)$ whose root is colored with $c_l$ in $S_{MNEB}(V)$ that does not belong to $Y_{[l],n-1}$, from theorem 1 we know that in $S_{MNEB}(e_{[l],r},V_r)$, either at most $k_l-2$ MNEBs belong to $Y_{[l],n-2}$, or there exists $h\ne l$, $1\leq h \leq R$, that at most $k_h-1$ MNEBs belong to $Y_{[h],n-2}$. Therefore, in $S_{MNEB}(V_r)$, there exists $1\leq h \leq R$, that at most $k_h-1$ MNEBs belong to $Y_{[h],n-2}$. From the induction hypothesis, we know that $V_r \notin S_{n-1}$, that means after the $(n-1)^{th}$ pruning, in the $l^{th}$ layer, at most $k_l-1$ neighbors of $V$ survived. So either $V$ has been pruned in the $(n-1)^{th}$ or even before, or it survived in the $(n-1)^{th}$ pruning but would be deleted in the $n^{th}$ pruning since its remaining neighbors in the $l^{th}$ layer are less than $k_l$ after the $(n-1)^{th}$ pruning, then we have $V\notin S_n$.

At this point, the sufficiency and necessity are proved, and  Theorem 2 can be established. \\

\noindent
%
%
%

\begin{figure}[!h]
\centering
\includegraphics[width=8cm]{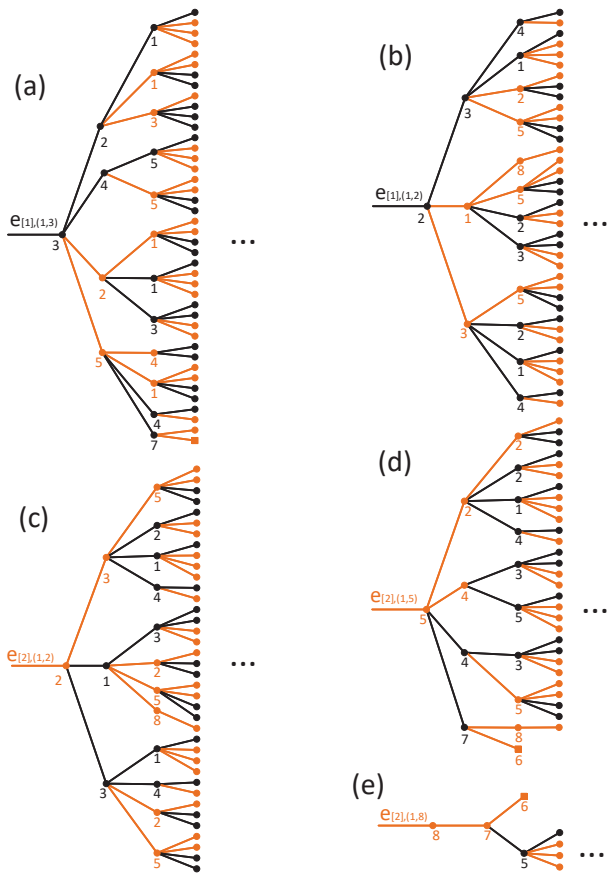}
\caption{
All $5$ MNEBs in $S_{MNEB}(1)$ of the network shown in Fig.~\ref{multilayer-fig1}. 
}
\label{multilayer-fig3}
\end{figure}

\textbf{An example of Theorem 2}:

Fig.~\ref{multilayer-fig3}\textbf{(a)}-\textbf{(e)} are $B(e_{[1],(1,3)},3)$, $B(e_{[1],(1,2)},2)$, $B(e_{[2],(1,2)},2)$, $B(e_{[2],(1,5)},5)$, $B(e_{[2],(1,8)},8)$, respectively. For $\mathbf{k}=(2,1)$ core decomposition, we can find that $B(e_{[1],(1,3)},3)\in Y_{[1],\infty}$, $B(e_{[1],(1,2)},2)\in Y_{[1],\infty}$, $B(e_{[2],(1,2)},2)\in Y_{[2],\infty}$, $B(e_{[2],(1,5)},5)\in Y_{[2],\infty}$, $B(e_{[2],(1,8)},8)\notin Y_{[2],1}$. So in $S_{MNEB}(1)$, there are two MNEBs that belong to $Y_{[1],\infty}$ and two MNEB that belong to $Y_{[2],\infty}$. So we have vertex $1\in S_{\infty}$. For $\mathbf{k}=(2,2)$, we can find that $B(e_{[1],(1,3)},3)\in Y_{[1],2}$, but not belongs to $Y_{[1],3}$. $B(e_{[1],(1,2)},2)\in Y_{[1],3}$, but not belongs to $Y_{[1],4}$. $B(e_{[2],(1,2)},2)\in Y_{[2],3}$, but not belongs to $Y_{[2],4}$. $B(e_{[2],(1,5)},5)\in Y_{[2],1}$, but not belongs to $Y_{[1],2}$. $B(e_{[2],(1,8)},8)\notin Y_{[2],1}$. So we can conclude that the vertex $1$ survives in the first two steps but will be deleted in the third pruning step.

\nocite{*}

\bibliography{bibliography}

\end{document}